\documentclass[12pt]{article}
\usepackage[english,german,french,polish]{babel}
\usepackage[T1]{fontenc}
\usepackage{amsfonts}

\begin{document}

\selectlanguage{english}

\baselineskip 0.75cm
\topmargin -0.6in
\oddsidemargin -0.1in

\let\ni=\noindent

\renewcommand{\thefootnote}{\fnsymbol{footnote}}

\pagestyle {plain}

\setcounter{page}{1}

\pagestyle{empty}

~~~

\begin{flushright}
IFT-- 03/24
\end{flushright}

{\large\centerline{\bf A proposal of neutrino mass formula{\footnote {Work supported in part by the Polish State Committee for Scientific Research (KBN), grant 2 P03B 129 24 (2003--2004).}}}}

\vspace{0.4cm}

{\centerline {\sc Wojciech Kr\'{o}likowski}}

\vspace{0.3cm}

{\centerline {\it Institute of Theoretical Physics, Warsaw University}}

{\centerline {\it Ho\.{z}a 69,~~PL--00--681 Warszawa, ~Poland}}

\vspace{0.3cm}

{\centerline{\bf Abstract}}

\vspace{0.2cm}

An explicit mass formula is proposed for three active neutrinos and their heavy sterile 
(righthanded) counterparts in the framework of seesaw mechanism. The formula correlates
reasonably the experimental estimates for the solar $\Delta m^2_{12}$ and atmospheric
$\Delta m^2_{32}$ in the case of hierarchical spectrum $m^2_1 \ll m^2_2 \ll m^2_3 $ of active 
neutrinos. The lightest heavy neutrino is predicted to be $O(10^5)$ times lighter than the heaviest.
An efficient mass formula, proposed previously for charged leptons, is recalled and coordinated 
with the new mass formula for neutrinos.

\vspace{0.2cm}

\ni PACS numbers: 12.15.Ff , 14.60.Pq .

\vspace{0.6cm}

\ni September 2003  

\vfill\eject

~~~
\pagestyle {plain}

\setcounter{page}{1}

As is well known, the bilarge form of neutrino mixing matrix

\begin{equation}
U = \left( \begin{array}{ccc} c_{12} & s_{12} & 0 \\ - \frac{1}{\sqrt2} s_{12} & \frac{1}{\sqrt2} c_{12} & \frac{1}{\sqrt2}  \\ \frac{1}{\sqrt2} s_{12} & -\frac{1}{\sqrt2} c_{12} & \frac{1}{\sqrt2}  \end{array} \right)\;, 
\end{equation}

\ni where $c_{12} = \cos \theta_{12}$ and $s_{12} = \sin \theta_{12}$ are estimated to correspond to $\theta_{12} \sim 33^\circ $,  is globally consistent with all present neutrino-oscillation experiments [1] for solar $\nu_e$'s and atmospheric $\nu_\mu$'s as well as KamLAND and Chooz  reactor $\bar\nu_e$'s [the negative result of the Chooz experiment gives the upper limit $s^2_{13} < 0.03$ for $s_{13} = \sin \theta_{13}$ that is neglected in (1)]. However, this form cannot explain the possible LSND effect [2] for accelerator $\bar\nu_\mu $'s (and $\nu_\mu $'s) whose existence (for $\nu_\mu $'s) is expected to be clarified soon in the miniBooNE experiment. Its negative result, consistent with the form (1) of $U$, would exclude mixings of active neutrinos with hypothetical light sterile neutrinos, leaving us with the 3 to 3 mixing transformation

\begin{equation}
\nu_{\alpha}  = \sum_i U_{\alpha i}  \nu_i 
\end{equation}

\ni between flavor and mass active neutrinos, $\nu_\alpha = \nu_e \,,\, \nu_\mu \,,\, \nu_\tau$ and $\nu_i = \nu_1\,,\,\nu_2\,,\,\nu_3$. Here, $U = \left(U_{\alpha i} \right)$ is given as in Eq. (1).  

In this note we make the observation that in the neutrino mass spectrum as it is seen in the present oscillation experiments there may be some new numerical regularities. Even if not understood theoretically yet, they may lead to a deeper insight in the neutrino mass problem. Needless to say that, in the history of physics, observations of new numerical regularities in energy or mass spectra of physical states often implied theoretical discoveries (a famous example is the empirical Balmer formula for hydrogen spectrum). 

We start with the neutrino $6\times 6$ mass matrix

\begin{equation}
\left( \begin{array}{cc} 0 & M^{(D)} \\ M^{(D)\,T} & M^{(R)} \end{array} \right) 
\end{equation}

\ni involving Dirac and righthanded Majorana $3\times 3$ mass matrices, $M^{(D)}$ and $M^{(R)}$, and accept the seesaw mechanism [3] leading to the effective $3\times 3$ mass matrix for active neutrinos of the form

\begin{equation}
M= M^{(D)} \frac{1}{M^{(R)}} M^{(D)\,T} \,,
\end{equation}

\ni where $M^{(R)}$ is assumed to dominate over $M^{(D)}$. Here,

\begin{equation}
U^T M U =  {\rm diag} (m_1, m_2, m_3) 
\end{equation}

\ni with $m_i$ denoting masses of active mass neutrinos $\nu_i $. For simplicity, we will take $M^{(D)}$ and $M^{(R)}$ real ($M^{(D)}$ is always Hermitian and $M^{(R)}$ --- symmetric), what is consistent with the real form (1) of $U$ (then, $U^T = U^\dagger = U^{-1}$).

Suppose that the spectrum of $M^{(D)}$ is identical with the masses $ m_{e_i} = m_e , m_\mu , m_\tau $ of charged leptons $e_i = e^- , \mu^- , \tau^- $ (though $M^{(D)}$ is the Dirac mass matrix for {\it neutrinos}), and assume that

\begin{equation}
U^\dagger M^{(D)} U = {\rm diag} (m_e , m_\mu , m_\tau) \,.
\end{equation}

\ni Then, from Eqs. (4), (5) and (6) we get for the active mass neutrinos $\nu_i$ the mass spectrum

\begin{equation}
m_i = \frac{m^2_{e_i}}{M_i}\,,
\end{equation}

\ni where 

\begin{equation}
\frac{1}{M_i} \equiv \sum_{\alpha \beta} U^*_{\alpha i}\left(\frac{1}{M^{(R)}}\right)_{\alpha \beta} U_{\beta i} 
\end{equation}

\ni and $M_i \gg m_{e_i} \gg m_i$ (with nonnegative $m_i$). In particular, if also the matrix $U^\dagger M^{(R)} U $ is diagonal, then $1/M_i$ are eigenvalues of $1/M^{(R)}$, since in this case

\begin{equation}
U^\dagger M^{(R)} U = {\rm diag} (M_1 , M_2 , M_3) \,,
\end{equation}

\ni where $M_i$ become Majorana masses of heavy sterile (righthanded) neutrinos. In general, however, $1/M_i$ given in Eq. (8) are average values of $1/M^{(R)}$ in the mass states of active neutrinos.

Now, we can observe that the conjecture of the ({\it weighted}) proportionality between $M_i$ and $m_{e_i}$, namely

\begin{equation}
M_i \propto N^2_i m_{e_i} \;,\; N_i = 1,3,5  \,,
\end{equation}

\ni leads to the values of $\Delta m^2_{21} \equiv m^2_2 - m^2_1$ and $\Delta m^2_{32} \equiv m^2_3 - m^2_2$ consistent with the experimental estimates [1]

\begin{equation}
(\Delta m^2_{21})^{\rm exp}  \sim 7\times 10^{-5}\;{\rm eV}^2 \;,\; (\Delta m^2_{32})^{\rm exp}  \sim 2.5\times 10^{-3}\;{\rm eV}^2 \;,
\end{equation}

\ni if the neutrino mass hierarchy $m^2_1 \ll m^2_2 \ll m^2_3 $ is realized. In fact, the seesaw spectrum (7) and the conjecture (10) imply the ({\it weighted}) proportionality between $ m_{e_i}$ and $m_i$:

\begin{equation}
m_{e_i} \propto N^2_i m_i \;.
\end{equation}

\ni This gives

\begin{equation}
\frac{m_2}{ m_3} = \frac{25}{9} \frac{m_\mu}{ m_\tau} = 0.1652 \;,
\end{equation}

\ni where the experimental values $ m_\mu = 105.658$ MeV and $ m_\tau = 1776.99^{+0.29}_{-0.26}$ MeV [4] are used, while for the experimental estimates $m^{\rm exp}_2 \sim \sqrt{7\times 10^{-5}}$ eV and $m^{\rm exp}_3 \sim \sqrt{2.5\times 10^{-3}}$ eV [valid if $m^2_1 \ll m^2_2 \ll m^2_3 $ in Eqs. (11)] one obtains 

\begin{equation}
\frac{m^{\rm exp}_2}{m^{\rm exp}_3} \sim \sqrt{2.8\times 10^{-2}} = 0.17
\end{equation}

\ni (for $m^{\rm exp}_3 \sim \sqrt{2\times 10^{-3}}$ eV this ratio is equal to 0.19). In another way, Eq. (13) {\it predicts} either

\begin{equation}
m_3 \sim \sqrt{2.6\times 10^{-3}}\, {\rm eV } = 5.1\times 10^{-2}\; {\rm eV}
\end{equation}

\ni with the input $ m_2 = m^{\rm exp}_2  \sim \sqrt{7\times 10^{-5}}$ eV or 

\begin{equation}
m_2 \sim \sqrt{6.8\times 10^{-5}}\, {\rm eV} = 8.3\times 10^{-3}\, {\rm eV}
\end{equation}

\ni with the input $ m_3 = m^{\rm exp}_3 \sim \sqrt{2.5\times 10^{-3}}$ eV (for $m_3 = m^{\rm exp}_3 \sim \sqrt{2\times 10^{-3}}$ eV the second prediction becomes $m_2 \sim \sqrt{5.5\times 10^{-5}}$ eV = $7.4\times 10^{-3}$ eV). Note that $m^2_2 \ll m^2_3$, in consistency with the assumption of neutrino mass hierarchy. 

From Eq. (12) we can also estimate $m_1$. In fact, Eq. (12) gives

\begin{equation}
\frac{m_1}{ m_2} = 9 \frac{m_e}{ m_\mu} = 0.0435271 \;, 
\end{equation}

\ni where the experimental values $m_e = 0.510999$ MeV and $m_\mu = 105.658$ MeV [4] are used, what leads to the {\it prediction}

\begin{equation}
m_1 \sim \sqrt{1.3\times 10^{-7}}\;{\rm eV} = 3.6\times 10^{-4}\; {\rm eV} \;,
\end{equation}

\ni when the input $m_2 = m^{\rm exp}_2 \sim \sqrt{7\times 10^{-5}}$ eV is applied. Notice that consistently $m^2_1 \ll m^2_2$. So, $m_1$ is $O(10^2)$ times smaller than $m_3$.

In conclusion, we can see that, in the framework of seesaw mechanism, our conjecture (10) is consistent with the present neutrino-oscillation data.

The proportionality coefficient for Eq. (12), call it $\zeta$, can be determined directly. {\it E.g.} taking $m_2 = m^{\rm exp}_2 \sim \sqrt{7\times 10^{-5}}$ eV and $m_\mu = 105.658$ MeV we calculate

\begin{equation}
\zeta = \frac{m_\mu}{9 m_2} \sim 1.4\times 10^9\;.
\end{equation}

\ni With the proportionality coefficient $\zeta $, Eq. (12) gives

\begin{equation}
m_{e_i}  = \zeta N^2_i m_i \;.
\end{equation}

\ni Then, from Eqs. (7) and (20)

\begin{equation}
M_i = \frac{m^2_{e_i}}{m_i }= \zeta^2 N^4_i m_i = \zeta N^2_i m_{e_i} \;.
\end{equation}

\ni Thus, $\zeta$ is the proportionality coefficient also for the conjecture (10). With Eq. (19), the formula (21) implies the estimates

\begin{equation}
M_1 \sim 7.2\times 10^5\; {\rm GeV}\,,\, M_2 \sim 1.3\times 10^9\; {\rm GeV}\,,\, M_3 \sim 6.2\times 10^{10}\; {\rm GeV} \;.
\end{equation}

\ni Hence, $M_1 \;{\bf :}\; M_2 \;{\bf :}\; M_3 = 1 \;{\bf :}\; 1.9\times 10^3 \;{\bf :}\; 8.7\times 10^4 $ and so, $M_1$ is predicted to be $O(10^5)$ times smaller than $M_3$.

Some time ago we observed that the mass spectrum of charged leptons $e_i = e^- , \mu^- , \tau^- $ can be expressed with high precision by the formula [5]

\begin{equation}
m_{e_i} = \rho_i \mu \left(N^2_i + \frac{\varepsilon -1}{N^2_i} \right) \,,
\end{equation}

\ni where 

\begin{equation}
\rho_i = \frac{1}{29} \,,\,\frac{4}{29} \,,\,\frac{24}{29} 
\end{equation}

\ni ($\sum_i \rho_i = 1$), $N_i = 1,3,5$ as above, and $\mu > 0$ and $\varepsilon > 0$ are constants. Then, we get explicitly

\begin{equation}
m_e = \frac{\mu}{29} \varepsilon \;,\;  m_\mu = \frac{\mu}{29} \frac{4}{9} (80+\varepsilon) \;,\;,  m_\tau = \frac{\mu}{29} \frac{24}{25} (624+\varepsilon)\;.
\end{equation}

\ni The interested reader may find a discussion about theoretical background of the simple formula (23) in Ref. [5] (in particular, the numbers $N_i$ and $\rho_i\; (i=1,2,3)$ are interpreted there). With the experimental values $m_e$ and $m_\mu$ as an input, the formula (23) leads to the {\it prediction}

\begin{equation}
m_\tau = \frac{6}{125}(351 m_\mu - 136 m_e) = 1776.80 \;{\rm MeV}
\end{equation}

\ni and also determines

\begin{equation}
\mu = \frac{29(9m_\mu - 4m_e)}{320} = 85.9924 \;{\rm MeV} \;,\; \varepsilon = \frac{320 m_e}{9m_\mu - 4m_e} = 0.172329 \,.
\end{equation}

\ni We can see that the prediction (26) lies really close to the experimental value $ m^{\rm exp}_\tau = 1776.99^{+0.29}_{-0.26}$ MeV. 

Making use of the charged-lepton mass formula (23) we can rewrite Eqs. (20) and (21) in the forms

\begin{equation}
m_i= \rho_i \frac{\mu}{\zeta} \left(1+ \frac{\varepsilon - 1}{N^4_i}\right) \,,\, M_i = \rho_i \,\mu \,\zeta N^4_i \!\left(1+ \frac{\varepsilon - 1}{N^4_i} \right) ,
\end{equation}

\ni where all factors on the rhs are known. The first and second Eq. (28) gives the mass formula for active and heavy sterile (righthanded) neutrinos, respectively. With the estimate (19) for $\zeta $ (determined by $m_2 \sim \sqrt{7\times 10^{-5}}$ eV), we get from Eqs. (28) the estimations (18) and (15) for $m_1$ and $m_3$ as well as the estimation (22) for $M_1$, $M_2$ and $M_3$. Notice that here the simple sum rule

\begin{equation}
\zeta N^2_i m_i + \frac{1}{\zeta N^2_i}M_i = 2m_{e_i}
\end{equation}

\ni ($i = 1,2,3$) holds.

\vfill\eject

~~~~
\vspace{0.5cm}

{\centerline{\bf References}}

\vspace{0.5cm}

{\everypar={\hangindent=0.6truecm}
\parindent=0pt\frenchspacing

{\everypar={\hangindent=0.6truecm}
\parindent=0pt\frenchspacing

[1]~For a recent review {\it cf.} V. Barger, D. Marfatia and K. Whisnant, {\tt hep--ph/0308123}.

\vspace{0.2cm}

[2]~For a report {\it cf.} G. Mills, {\it Nucl. Phys. Proc. Suppl.} {\bf 91}, 198 (2001); and references therein.

\vspace{0.2cm}

[3]~M. Gell-Mann, P. Ramond and R.~Slansky, in {\it Supergravity}, edited by F.~van Nieuwenhuizen and D.~Freedman, North Holland, 1979; T.~Yanagida, Proc. of the {\it Workshop on Unified Theory and the Baryon Number in the Universe}, KEK, Japan, 1979; R.N.~Mohapatra and G.~Senjanovi\'{c}, {\it Phys. Rev. Lett.} {\bf 44}, 912 (1980).

\vspace{0.2cm}

[4]~The Particle Data Group, {\it Phys.~Rev} {\bf D 66}, 010001 (2002).

\vspace{0.2cm}

[5]~W. Kr\'{o}likowski, {\it Acta Phys. Pol.} {\bf B 33}, 2559 (2002); and references therein. 

\vfill\eject

\end{document}